\documentclass[runningheads]{llncs}
\usepackage{graphicx}
\usepackage{gensymb}
\usepackage{amsmath}
\usepackage[mathletters]{ucs}
\usepackage[utf8x]{inputenc}
\usepackage{aasmacros}
\usepackage{booktabs}
\usepackage[tableposition=above]{caption}
\usepackage[svgnames,table,dvipsnames]{xcolor}
\usepackage{marvosym}
\definecolor{darkgreen}{rgb}{0.05, 0.5, 0.1}
\definecolor{facu}{rgb}{0.05, 0.1, 0.5}
\begin{document}

\title{Reddening-free $Q$ indices to identify Be star candidates}

\author{Yael Aidelman\inst{1,2}\orcidID{0000-0001-5279-0241} \and
Carlos Escudero\inst{2}\orcidID{0000-0002-6056-6247} \and
Franco Ronchetti\inst{3,4}\orcidID{0000-0003-3173-1327} \and
Facundo Quiroga\inst{3}\orcidID{0000-0003-4495-4327}\Letter \and
Laura Lanzarini\inst{3}\orcidID{0000-0001-7027-7564} }
\authorrunning{Y. Aidelman et al.}
%
\institute{
Departamento de Espectroscop\'{\i}a, Facultad de Ciencias Astron\'omicas y Geof\'{\i}sicas, Universidad Nacional de La Plata (UNLP), Paseo del Bosque S/N, B1900FWA, La Plata, Argentina.\and
Instituto de Astrof\'{\i}sica La Plata, CCT La Plata, CONICET-UNLP, Paseo del Bosque S/N, B1900FWA, La Plata, Argentina. \and
Instituto de Investigación en Informática LIDI, Facultad de Informática, Universidad Nacional de La Plata, La Plata, Argentina
\and
Comisión de Investigaciones Científicas de la Pcia. De Bs. As. (CIC-PBA) }
\maketitle              

\begin{abstract}

Astronomical databases currently provide high-volume spectroscopic and photometric data. While spectroscopic data is better suited to the analysis of many astronomical objects, photometric data is relatively easier to obtain due to shorter telescope usage time. Therefore, there is a growing need to use photometric information to automatically identify objects for further detailed studies, specially H$\alpha$ emission line stars such as Be stars.
Photometric color-color diagrams (CCDs) are commonly used to identify this kind of objects.
However, their identification in CCDs is further complicated by the reddening effect caused by both the circumstellar and interstellar gas. This effect prevents the generalization of candidate identification systems. 
Therefore, in this work we evaluate the use of neural networks to identify Be star candidates from a set of OB-type stars. The networks are trained using a labeled subset of the 
VPHAS+ and 2MASS databases, with filters $u, g, r,$ H$α, i, J, H$, and $K$. In order to avoid the reddening effect, we propose and evaluate the use of reddening-free $Q$ indices to enhance the generalization of the model to other databases and objects. To test the validity of the approach, we manually labeled a subset of the database, and use it to evaluate candidate identification models. We also labeled an independent dataset for cross dataset evaluation. We evaluate the recall of the models  at a 99\% precision level on both test sets. Our results show that the proposed features provide a significant improvement over the original filter magnitudes.

%

\end{abstract}

\section{Introduction}
\label{intro}




In the big data era, free access to
databases in different wavelength ranges, from gamma-rays to radio waves, together with machine-learning methods, has drastically incremented
 the possibility to study and identify different types of peculiar line-emission stars using photometric information (e.g., Vioque et al., 2019~\cite{Vioque2019}; Akras et al., 2019~\cite{Akras2019}; Pérez-Ortiz et al., 2017~\cite{Perez-Ortiz2017}).

While spectroscopic techniques are excellent to perform accurate stellar classification and deepen into the study of various spectral features, the telescope time required to obtain such information is longer compared to obtaining photometric data.


The goal of this work is then to use the potential of photometric data to search for emission-line star candidates. These can be later observed and be confirmed spectroscopically as such. Particularly, we are interested in detecting Be star candidates.

Be stars are emission-line objects that rotate at high speed (Jaschek et al., 1981~\cite{Jaschek1981}; Struve, O., 1931~\cite{Struve1931}) and constitute unique astrophysical laboratories. They are of interest in various branches of stellar physics dedicated to the study of mechanisms of mass loss, angular momentum distribution, astroseismology, among others.

The rest of this section describes Be stars in detail, classical techniques to detect plausible candidates as well as previous star candidate proposals based on machine-learning methods.

\subsection{Be stars}

Be stars are defined as non-supergiant spectral B-type stars that exhibit, or have exhibited, one or more hydrogen lines in emission (Jaschek et al., 1981 ~\cite{Jaschek1981}; Collins, II, G., 1987~\cite{Collins1987}), particularly the H$\alpha$ line.
In some cases, it is also possible to observe the presence of once-ionized helium and metal lines in emission.  Thus, this definition not only applies to B-type stars but also to late O- and early A-type stars.

The analysis of spectrophotometric observations of Be stars at different wavelengths, combined with interferometric and polarimetric data (Gies et al., 2007~\cite{Gies2007}; Meilland et al., 2007~\cite{Meilland2007}, among  others), 
indicate that the different properties shown by these stars could be interpreted by the existence of an optically-thin gaseous circumstellar equatorial disk in Keplerian motion (see Rivinius et al., 2013~\cite{Rivinius2013}). 
This suggests that the high rotation speed would play a significant role in the development of the equatorial disk
(e.g., Struve, O., 1931~\cite{Struve1931}; Huang, S., 1972~\cite{Huang1972}; Quirrenbach, A., 1993~\cite{Quirrenbach1993}; Quirrenbach et al., 1994~\cite{Quirrenbach1994}; Hirata, R., 1995~\cite{Hirata1995}).
However, despite  the increasing observational evidence that Be stars do not rotate at their critical rotational speed (Zorec et al. 2016~\cite{Zorec2016}, Zorec et al. 2017~\cite{Zorec2017}, Aidelman et al. 2018~\cite{Aidelman2018}, Cochetti et al., 2019~\cite{Cochetti2019}), 
there is still no consensus on disk formation mechanism(s).

Other observed effects induced by stellar rotation during the main sequence phase of hot stars, are the development of axi-symmetric winds, the modification in pulsation modes, changes in metallicity or the presence of magnetic fields (see Peters et al., 2020~\cite{Peters2020}; Rivinius et al., 2013~\cite{Rivinius2013}).
These  properties make Be stars perfect stellar laboratories, of interest in different astrophysical topics, as mentioned above.

In this context, the discovery, classification and analysis of a considerable sample of Be stars in different environments are necessary to understand their nature.

\subsection{Related work}

 To the best of our knowledge, there are no previous works focused on Be stars that use the reddening-free $Q$ indices as we propose in this work (see section~\ref{features}). Therefore we briefly summarize work similar to ours.

Pérez-Ortiz et al. (2017~\cite{Perez-Ortiz2017}) select Be star candidates using light curves of the $I$ band obtained from OGLE-IV data (Udalski et al., 2015~\cite{udalski2015ogle}). They train classification trees, random forest, gradient boosted trees, support vector machines (SVM) and K-nearest neighbours on OGLE-III data. To evaluate the models, they compare the average f-score from 10-fold cross validation. While random forests achieve the best scores, most models behave similarly. To improve the cross-dataset robustness of their models, they employ a custom feature based on fourier coefficients of the data. They propose 50 new Be star candidates selected from OGLE-IV data.

Vioque et al., (2019~\cite{Vioque2019}) use photometric data similar to ours which includes a passband filter for the H$\alpha$ wavelength. However, their sources include many other filters, resulting in 48 variables for each sample. In order to avoid problems caused by interstellar extinction, they select objects for which the effect of this phenomena is negligible. They apply  Principal Component Analysis (PCA) to reduce the features to a 12 latent dimensions which contain 99.99\% of variability. Afterwards, they employ a neural network composed of three linear layers to classify candidates.

Akras et al., (2019~\cite{Akras2019}) identify symbiotic stars from other objects. Their data includes various photometric filters that can detect H$\alpha$. They perform a thorough manual evaluation of color-color\footnote{A color index is defined as the difference of two magnitudes at different wavelengths ($m_{\lambda_1} - m_{\lambda_2}$). Magnitude is a unitless measure of the brightness of an object on a logarithmic scale in a defined passband. The brighter an object, the more negative the value of its magnitude} diagrams (CCDs) to identify feature combinations which can separate these stars from other kinds of similar objects. Afterwards, they repeat this approach to classify symbiotic stars into subsets. Their approach can identify a small subset of previously labeled symbiotic stars and also proposes 125 new candidates. They employ a combination of k-nearest neighbours, linear discriminant analysis, and classification trees as models.

\subsection{Proposed work}

One technique commonly used to identify classical Be star candidates relies on photometric color-color diagrams. 
These diagrams use differences of apparent magnitudes between narrow-band filters, such as H$\alpha$ passband, and broad-band filters centered at other given wavelengths, such as the filters $r$ or $i$.
However, these colors are affected by interstellar extinction.
Therefore, in this work we propose to use the reddening-free $Q$ photometric indices, as described in section~\ref{Features}.

The construction of the $Q$ indices from different apparent magnitudes opens up the potential of broad-band and narrow-band photometric data together with  machine-learning techniques to quickly obtain a significant number of Be star candidates observed in any direction in the sky.
Subsequently, this method will allow us to carry out a rapid and accurate spectroscopic follow-up of these stars to confirm their  classification and properties.

\section{Datasets and Features}
\subsection{Datasets}
\label{datasets}

\begin{figure}
\centering
\includegraphics[width=0.9\textwidth]{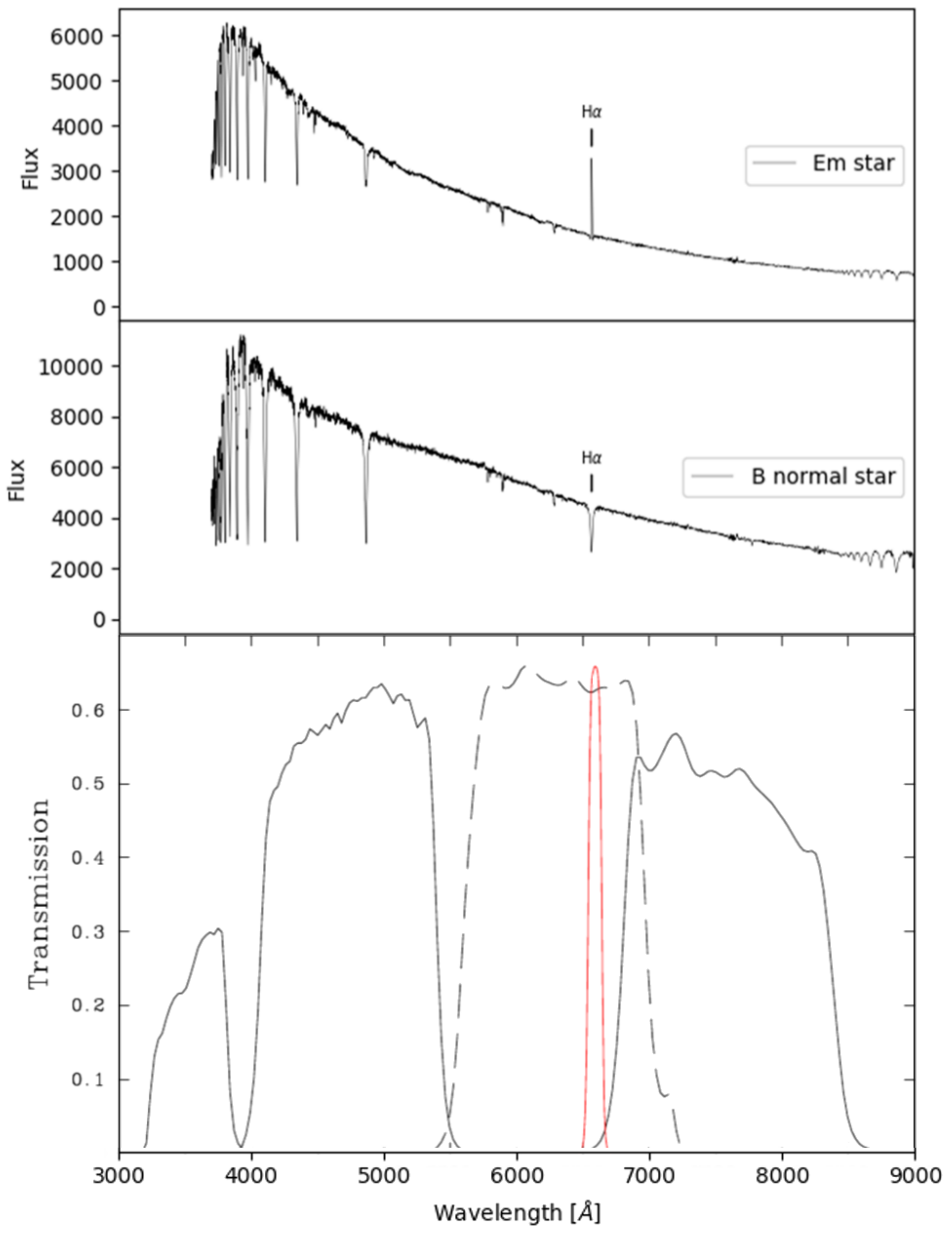}
\caption{Upper and middle panel show two B-type star spectra from LAMOST (Cui et al., 2012~\cite{Cui2012}). The upper panel corresponds to a Be star, while the middle panel to a typical B-type star. Bottom panel shows the transmission profiles of the filters used by VPHAS+ (Drew et al., 2014~\cite{Drew2014}).
The H$\alpha$ narrow-band filter is shown in red.
} \label{spectra}
\end{figure}

We use the data published by Mohr-Smith et al. (2017~\cite{Mohr-Smith2017}) on the Carina Arm region ($282\degree \leq l \leq 293\degree$). These authors used data from the VST Photometric H$\alpha$ Survey of the Southern Galactic Plane and Bulge (VPHAS+; Drew et al., 2014~\cite{Drew2014}) in $u, g, r$, H$\alpha$, $i$ filters (see Figure~\ref{spectra} bottom panel) combined with $J, H, K$ magnitudes from the Two Micron All Sky Survey (2MASS; Skrutskie et al., 2006~\cite{Skrutskie2006}).
Performing fittings to the spectral energy distribution, Mohr-Smith et al. grouped the sample of OB-type stars in four groups: emission-line stars (EM), sub- and over-luminous stars, and normal stars. 
The features of the dataset consist of the fluxes at the 8 filters mentioned above. The dataset contains a sample of $5877$ OB-type stars labeled with four classes. Since we are only interested in detecting emission-line stars from the other classes, we group all non-EM stars into a single set called \emph{Normal OB}. The resulting class distribution is shown in Table~\ref{class_distribution}.



To test the inter-database accuracy of the model, we manually labeled a subset of OB-type stars classified spectroscopically by Liu et al. (2019~\cite{Liu2019}).
The photometric data of this subset was obtained from VPHAS+ dr2 (for the southern hemisphere). Data in filters $r$, H$\alpha, i$ were from the INT Photometric H$α$ Survey of the Northern Galactic Plane (IPHAS dr2; Barentsen et al., 2014~\cite{barentsen2014second}) while  the data in filters $u, g$ were obtained from the Sloan Digital Sky Survey (SDSS dr12; Alam et a., 2015~\cite{alam2015eleventh}), for the northern hemisphere.
Additionally,  the data in filters $J, H, K$ for both hemispheres were obtained from the 2MASS. Spectroscopic data is available from the Large Sky Area Multi-Object Fiber Spectroscopic Telescope (LAMOST dr5; Cui et al, 2012~\cite{Cui2012}).

We selected stars which have both photometric and spectroscopic data. Among the 22901 OB-stars classified by Liu, only 1113 have 
measurements for the same set of 8 filters used by Mohr-Smith et al.


Via visual inspection of the spectra, we were able to label 283 stars (among 1113 OB-stars with photometry). We identified OB-stars that present the H$\alpha$ line in emission (see Figure~\ref{spectra} middle and upper panel).
As shown in Table~\ref{class_distribution}, we identify 98 objects as EM stars.

\begin{table}
    \setlength{\tabcolsep}{8pt}
    \centering
    \caption{Class distribution of samples.}\label{class_distribution}
    \begin{tabular}{lccc}
        \toprule
        Dataset & Normal OB  & EM & \emph{Total}\\    
        \midrule
        Mohr-Smith et al. (2017~\cite{Mohr-Smith2017}) & 5629 & 248 & \emph{5877}\\
        Liu et al. (2019~\cite{Liu2019}) & {185} & {98} & {\emph{283} }\\
        \bottomrule
    \end{tabular}
\end{table}


\subsection{Features}
\label{features}
\label{Features}

As mentioned in Section \ref{datasets}, the data used corresponds to magnitudes 
(hereinafter original features) obtained in seven different broad-band filters: $u$, $g$, $r$, $i$, $J$, $H$, and $K$, and in one narrow-band filter, H$\alpha$.
However, the intrinsic magnitude of an object can be affected by several factors, such as the distance to the star and the interstellar extinction. Particularly, in the latter, the interstellar material (dust and gas) located between the observer and the object absorbs part of its radiation (mainly ultraviolet light). A priori, neither the amount of interstellar material in the visual direction nor the distance to the star are known. For these reasons, the use of magnitudes (available in the databases) as a tool to try to classify objects of different morphology or characteristics is not enough.

\begin{figure}
\includegraphics[width=0.505\textwidth]{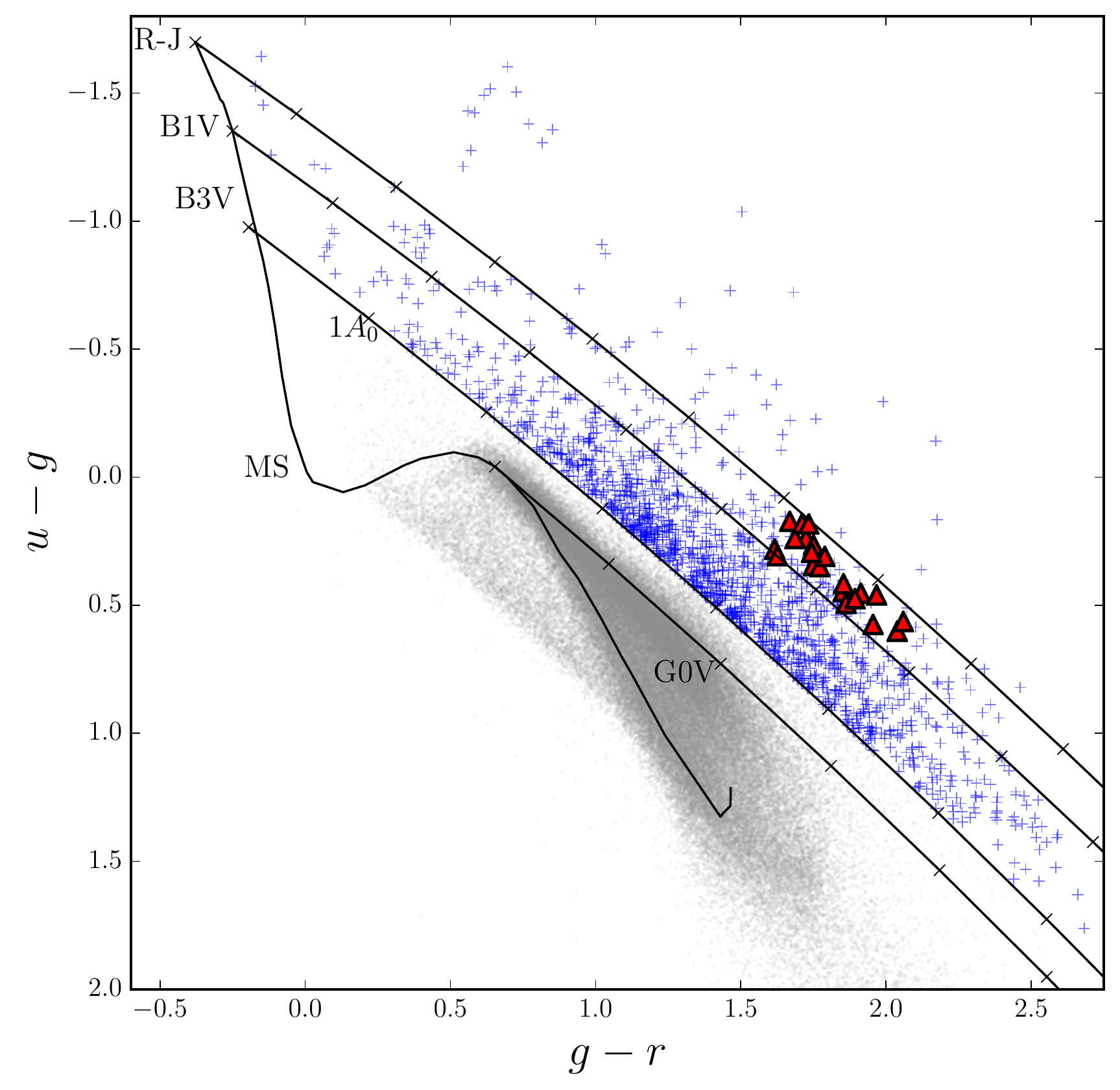}
\includegraphics[width=0.495\textwidth]{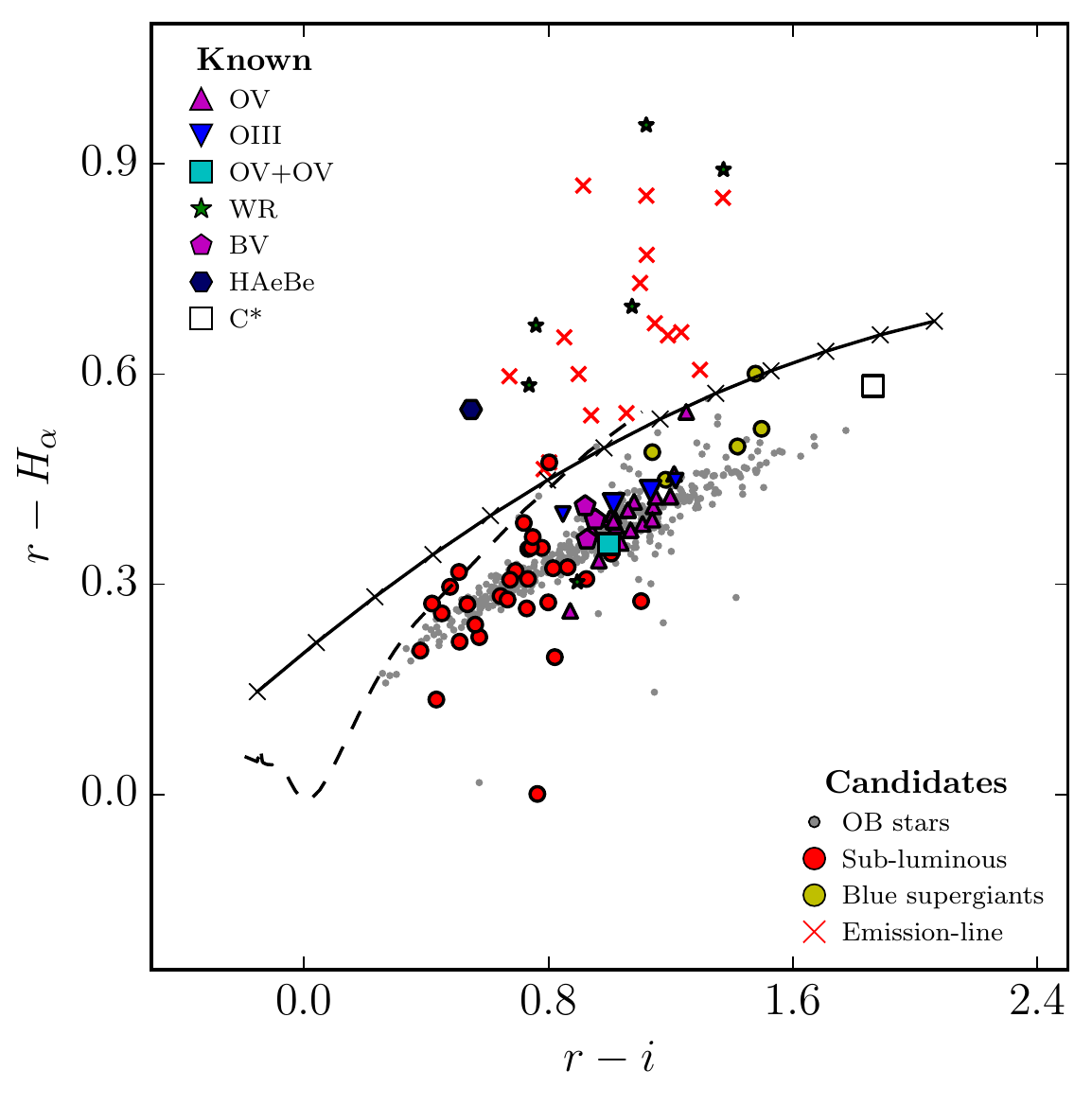}
\caption{Color-color diagrams taken from Mohr-Smith et al. (2015~\cite{Mohr-Smith2015}). Left: Location of B-type stars using $u,g,r$ apparent magnitudes (blue crosses), and their corresponding location in the main sequence (MS) if they were not
affected by interstellar extinction (black crosses). Right: location of the brightest stars in the H$\alpha$ passband.}
\label{CCD}
\end{figure}

On the other hand, although the color index is independent of distance, it is still affected by interstellar reddening, as shown in Figure~\ref{CCD}. The left panel shows the CCD done by Mohr-Smith et al. (2015~\cite{Mohr-Smith2015}) with the stars detected in the Carina Arm region (gray dots). The blue crosses represent B-type stars whose location in the diagram is
affected by the interstellar extinction.


In particular, as previously mentioned, one technique commonly used to identify classical Be star candidates is to use CCDs that combine a narrow-band filter centered at the H$\alpha$ line and a filter that samples the nearby continuum region. Figure \ref{CCD} (right panel) shows ($r-\mathrm{H}\alpha$) versus ($r-i$) diagram with the location of stars with different characteristics (Mohr-Smith et al., 2015~\cite{Mohr-Smith2015}). As seen in this figure, Be stars (red crosses) are separated from other objects because they show an excess emission in H$\alpha$. However, the presence of other astrophysical sources, such as Wolf Rayet (WR) and Herbig AeBe (HAeBe) stars, in the same region can still be observed.

For all the aforementioned, in order to avoid the effects of interstellar extinction and distance, we propose the use of the reddening-free $Q$ indices (hereinafter $Q$ features). This  index was introduced by Johnson \& Morgan, (1955~\cite{Johnson1955}) and it is defined as: 


\begin{equation}
\begin{split}
    Q_{1234} & = (m_{\lambda_1}-m_{\lambda_2}) - \dfrac{r_{\lambda_1}-r_{\lambda_2}}{r_{\lambda_3}-r_{\lambda_4}}\, (m_{\lambda_3}-m_{\lambda_4})\\
    & = (m_{\lambda_1}^{0}-m_{\lambda_2}^{0}) - \dfrac{r_{\lambda_1}-r_{\lambda_2}}{r_{\lambda_3}-r_{\lambda_4}}\, (m_{\lambda_3}^{0}-m_{\lambda_4}^{0})
\end{split}
\end{equation}

\noindent where $m_{\lambda_i}$ is the apparent (observed) magnitude and $m_{\lambda_i}^{0}$ is the apparent magnitude corrected by interstellar extinction effect
in four (in some cases three) different filters centered at given wavelength $\lambda_i$, $r_{\lambda_i} = A_{m_{\lambda_i}} / A_{V}$, were $A_{m_{\lambda_i}}$ and $A_{V}$ are the extinction coefficient for 
${\lambda_i}$ and for the $V$\footnote{Filter $V$ corresponds to Johnson's photometric system.} filter, respectively. The $r_{\lambda_i}$ values adopted in this work were those calculated by Schlafly et al. (2011~\cite{Schlafly2011}) for a selective absorption coefficient\footnote{
The selective absorption coefficient relates the absorption coefficient in the visual, $A_{\rm v}$, with the excess color $E(B-V)$, through the ratio $A_{\rm v} = R_{\rm v}\, E(B-V)$.} $R_{\rm V} = 3.1$.

\section{Experiments}
\label{experiments}
\subsection{Metodology}

In order to test the suitability of the $Q$ features, we trained and evaluated models with the original magnitudes and the features, separately. Additionally, we use Neighborhood Components Analysis (NCA; Goldberg et al., 2005~\cite{goldberger2005neighbourhood}) to perform dimensionality reduction to 2 components from the 56 variables of the $Q$ features (built with the combinations of the 8 original magnitudes taken from 3) and generate another set of features (see Table~\ref{tab:feature_list}). The 2-dimensional projection obtained by NCA can provide diagrams similar to the CCD commonly used in astronomy, albeit using latent variables (see section \ref{sec:results}).

\begin{table}
    \setlength{\tabcolsep}{8pt}
    \centering
    \caption{Features and their dimensionalities.}\label{tab:feature_list}
    \begin{tabular}{lp{17em}r}
        \toprule
        Features & Description & Dimensionality\\    
        \midrule
         Originals &  Magnitudes in $u$,$g$,$r$,H$\alpha$,$i$,$J$,$H$,$K$ filters & 8\\
        $Q$ & Reddening-free  index (section \ref{Features}) & 56 \\
        $NCA$ & Neighborhood Components Analysis$^*$ over $Q$ features & 2\\
        \bottomrule
    \end{tabular}
    \begin{flushleft}
    $^*$ Goldberg et al., 2005~\cite{goldberger2005neighbourhood}
    \end{flushleft}
\end{table}

We also compare different models in terms of their relative performances. We use a simple logistic regression as a baseline, and compare it against Support Vector Machines (SVM) with Linear and Gaussian kernels, with $C=10$ in both cases. We also compare Neural Networks with 1 hidden layer with $8$ linear units and $tanh$ activation function.

Given that candidate selection is essentially a binary classification task, the most natural overall performance metric is the F-score. However, since the goal of our work is to avoid manual verification of stars with a low probability of being Be stars, we focus on reducing the number of selected candidates. Therefore, we prefer to evaluate models in terms of their recall at a 99\% level of precision\footnote{We note that \emph{Purity} and \emph{Completeness} are commonly used as synonyms for \emph{Precision} and \emph{Recall}, respectively. These terms are more prevalent in astronomy.}. That is, for each model we set a threshold so that its precision is around 99\% and measure the resulting recall. 

All models are trained on the Mohr-Smith dataset (see Section \ref{datasets}). Given the small sample size of the datasets, we use random subsampling cross validation with $20$ random splits to obtain average values of each measure. We perform a 90/10 split, obtaining train/test sizes of approximately $5000$ and $500$ samples. For each split, a model is trained on Mohr-Smith, and evaluated on its test set. Afterwards, the same model is evaluated on the Liu dataset, using all samples as a test set. The same threshold used to obtain 99\% precision for Mohr-Smith is also employed when evaluating the Liu dataset.

\subsection{Results}
\label{sec:results}

Table \ref{tab:results} shows the recall rate for various model and feature combinations. The best result on the Liu dataset was obtained with Neural Network using $Q$ features, obtaining a 25\% recall rate. For the Mohr-Smith dataset, all models perform similarly.

Linear models, however, don't generalize as well as non-linear models for Liu. Nonetheless, a simple linear model with a non-linear dimensionality reduction such as NCA  (Figure \ref{fig:nca}) can obtain  almost 14\% recall in Liu. This indicates that the classes are not linearly separable even in the 56 dimensional space of the $Q$ features.

In the case of SVM, results are somewhat erratic (Figure \ref{tab:results}). This may be due to the fact that calibrating a SVM to output probabilities is usually difficult, given that the probability estimation model must be fitted on top of the SVM after the model is trained.

Given the recall increment when using $Q$ features ($+11.9$\%) from the best model using magnitudes (Gaussian SVM) to the best model using $Q$ features (Neural Network), we can conclude that these features indeed help with the identification of H$\alpha$ emitting stars.

\begin{table}
    \setlength{\tabcolsep}{8pt}
    \centering
    \caption{Recall of models on the Mohr-Smith and Liu datasets, with a threshold set for 99\% precision.}\label{tab:results}
    \begin{tabular}{llcc}
        \toprule
        Model & Features & Mohr-Smith (Recall)  & Liu (Recall) \\    
        \midrule
        Log. Regression & Magnitudes & 84.2 ($\pm$7)\% & 5.7 ($\pm$0.1)\% \\
        Log. Regression & $Q$ & 81.3 ($\pm$12)\% & 5.5 ($\pm$10)\% \\
        Log. Regression & NCA & 74.6($\pm$11)\% & 13.9 ($\pm$9)\% \\
        SVM (Linear) & Magnitudes & 82.4($\pm$14)\% & 0($\pm$0)\% \\
        SVM (Linear) & $Q$ & 85.2($\pm$9)\% & 9.2 ($\pm$11)\% \\
        SVM (Gaussian) & Magnitudes & 85.2($\pm$11)\% & 13.1 ($\pm$2)\% \\
        SVM (Gaussian) & $Q$ & 37.2($\pm$3)\% & 4.7 ($\pm$3)\% \\
        Neural Network & Magnitudes & 84.8($\pm$8)\% & 9.5($\pm$4)\% \\
        Neural Network & $Q$ & 85.2($\pm$14)\% & \textbf{25 ($\pm8$)\% }\\
        \bottomrule
    \end{tabular}
\end{table}

\begin{figure}
    \centering
    \includegraphics[width=0.86\textwidth]{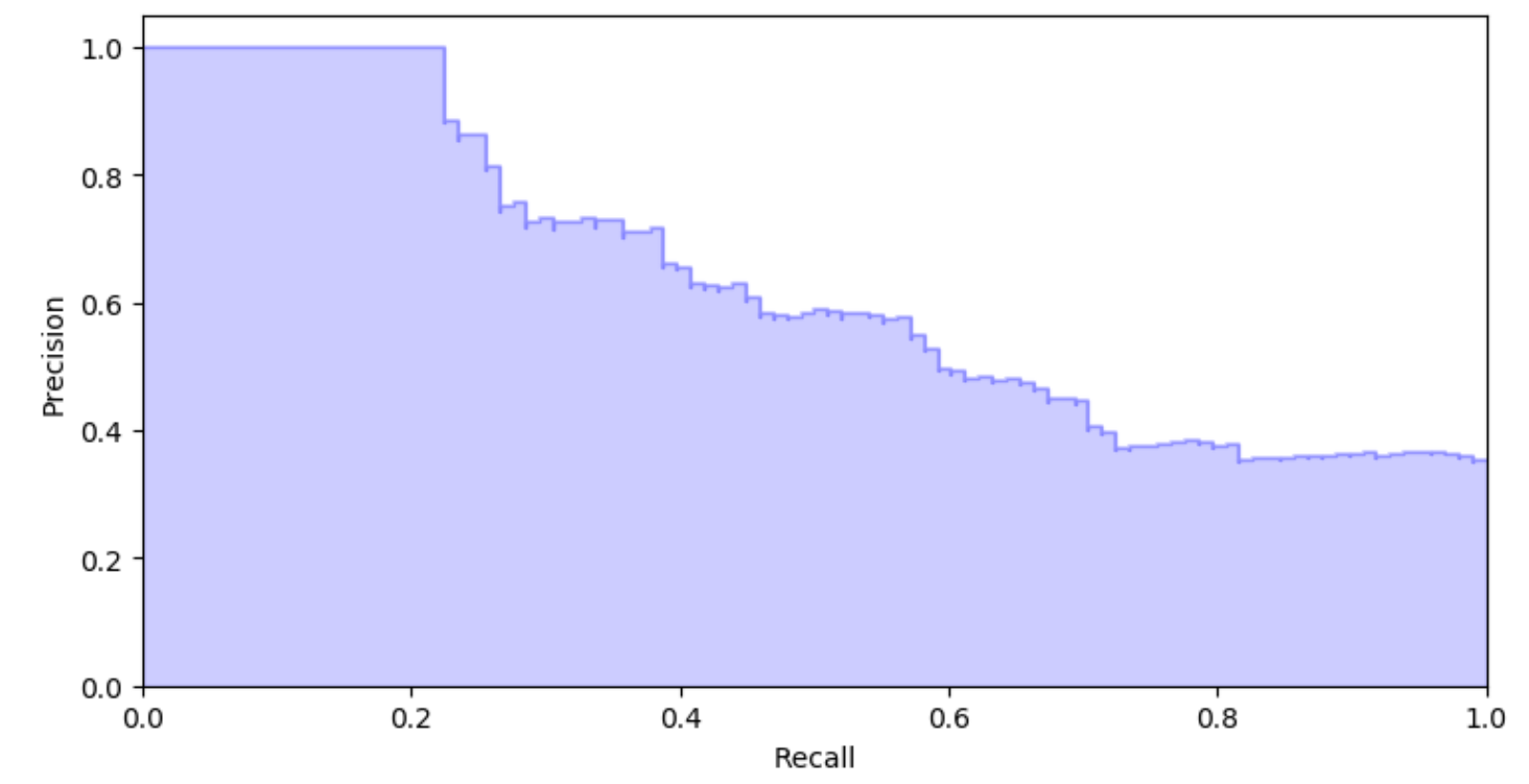}
    \caption{Precision-Recall curves in Liu dataset for a Neural Network trained with $Q$ features. }
    \label{fig:precision_recall}
\end{figure}

Figure \ref{fig:precision_recall} shows precision-recall curves evaluated in Liu dataset for a Neural Network model using the $Q$ features. As listed in Table \ref{tab:results}, the model achieves a 25\% recall rate with 100\% precision. For larger values of recall, precision rates drop in an approximately linear fashion, which indicates a good balance between these two metrics.

\subsection{Recall Rate Analysis}
\label{sec:analysis}

\begin{figure}
    \centering
    \includegraphics[width=0.99\textwidth]{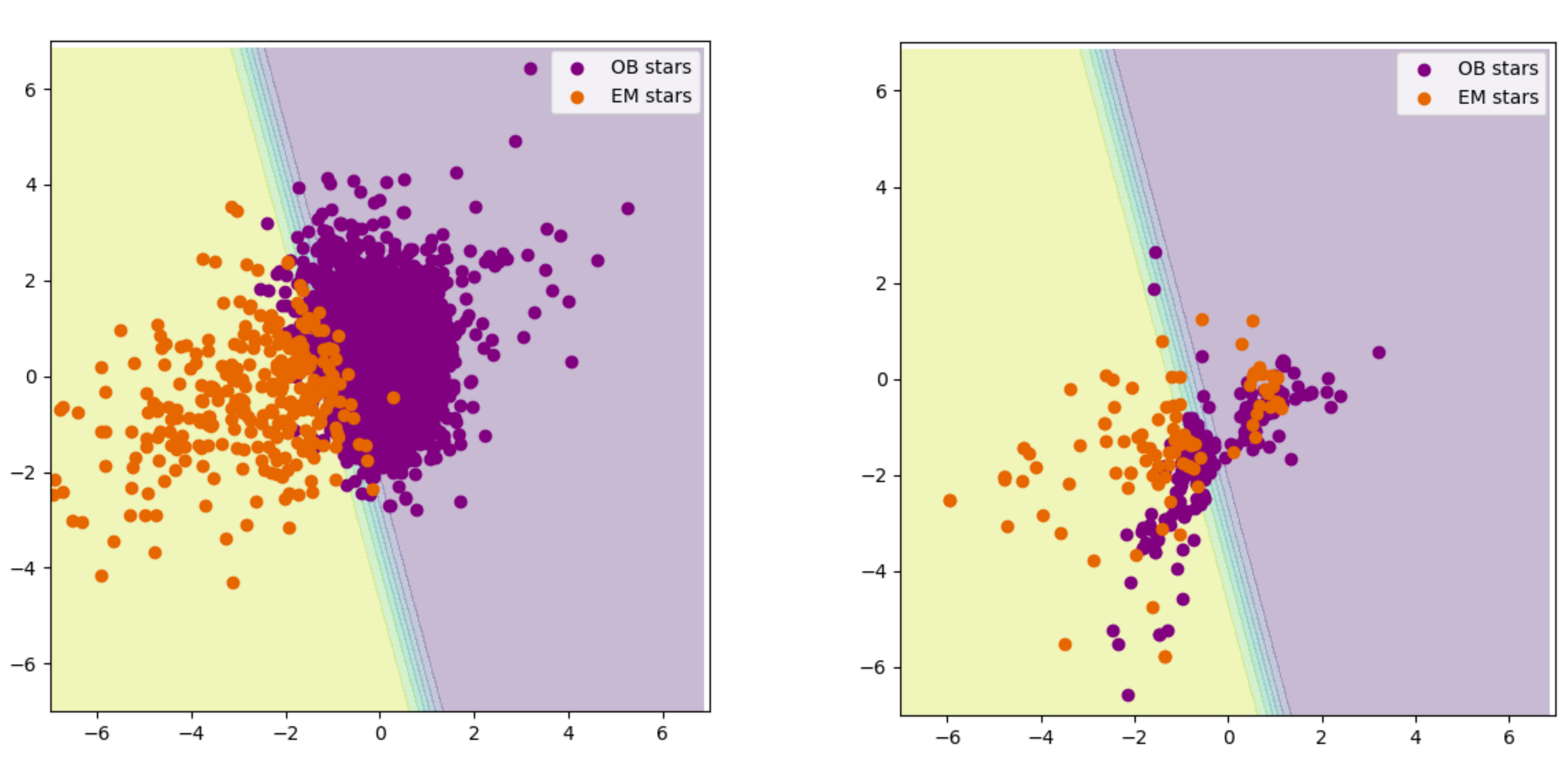}
    \caption{NCA features for the Mohr-Smith (left) and Liu (right) datasets. The decision lines corresponds to a Sigmoid function from a Logistic Regression classifier trained on Mohr-Smith.}
    \label{fig:nca}
\end{figure}

We plot the 2-dimensional NCA features for both test set of Mohr-Smith and Liu (Figure \ref{fig:nca}). As can be seen in the figure, a significant number of EM stars fall in the region of normal OB stars, and vice versa. For this reason, we decided to visually inspect the spectra of some objects in which this situation occurred. 

The high number of false-negative cases (reflected in the low recall rate) may be due to the detection limits existing in each observation technique.
In most of the objects classified as false negatives, the H$\alpha$ line profile is observed with the wings in emission (absorption) and the nucleus in absorption (emission) (Catanzaro, G., 2013~\cite{Catanzaro2013}; Dimitrov et al., 2018~\cite{Dimitrov2018}).
However, if the emission is not intense enough, it is not reflected in the photometry since the values of the color indices and $Q$ parameters correspond to that of a normal star.

On the other hand, false-negative and false-positive cases may also be due to the variability of the Be phenomenon. This effect is probably due to a significant increase in the circumstellar material that intensifies the emission in H$\alpha$. Subsequently, this intensity decreases with time and may disappear completely (Rivinius et al., 2013~\cite{Rivinius2013}; Dimitrov et al., 2018~\cite{Dimitrov2018}). 
Since the photometric and spectroscopic data are not simultaneous, it may happen that the photometric data does not reflect an emission in the H$\alpha$ passband even if the emission H$\alpha$ line in the spectrum is observed.
Therefore, for a correct photometric classification of these stars, it is necessary to carry out spectroscopic follow-up close in time between both observation modes.


\section{Conclusions and Future Work}
\label{conclusions}

In this work, we have compared different classification models to distinguish objects with emission in the H$\alpha$  line from normal OB stars. The models are trained with photometric data collected from various sources. 
Afterward, these models can be employed to identify Be star candidates, a type of H$\alpha$ emitting OB stars.

We also propose the use of the reddening-free $Q$ indices to remove unwanted variations caused by interstellar extinction effects.  Experiments show these features improve the cross-dataset performance of the classification models, obtaining up to 25\% recall for 99\% precision on an unseen dataset. Meanwhile, the best model trained with the original features obtains at most 13.1\% recall at the same precision level.

In order to test the cross-dataset performance of the models, we manually labeled samples obtained from different sources to form a new test set. Posterior analysis shows that there is significant variability in the cross-dataset test samples, which the model cannot learn given the original training data.

The presented models can detect H$\alpha$ emitting stars, which are good candidates for Be stars. However, confirmation requires spectroscopic follow up. We plan on expanding the pipeline and train classification models to also identify different classes of H$\alpha$ emitting objects, including Be stars, with a higher degree of automation.
We will also focus on expanding the data sources used, in order to both train a more robust model and test with more variable sets of objects.

\vspace{0.5cm}
{\it{\bf Acknowledgements}

This work is based on data obtained as part of the INT H$\alpha$ photometric study of the northern galactic plane (IPHAS; https://www.iphas.org), VST Photometric H$\alpha$ Survey of the Southern Galactic Plane and Bulge (VPHAS+; https://\-www.vphasplus.org),  Two Micron All Sky Survey (2MASS, https://irsa.ipac.\-caltech.edu/Missions/2mass.html), Sloan Digital Sky Survey (SDSS; https://\-www.sdss.org) and The Large Sky Area Multi-Object Fibre Spectroscopic Telescope (LAMOST; http://www.lamost.org). 

YA is grateful to L.~Cidale, G.~Baume and A.~Smith Castelli for their helpful comments and suggestions.
}

\bibliographystyle{splncs04}
\bibliography{mybibliography.bib}
\end{document}